\documentclass[eqsecnum,showpacs,showkeys,nofootinbib,aps]{revtex4}

\usepackage{epsfig}

\renewcommand{\theequation}{\arabic{equation}}
\def\be{\begin{equation}}
\def\ee{\end{equation}}
\def\bea{\begin{eqnarray}}
\def\eea{\end{eqnarray}}

\def\na{\nabla}

\begin{document}

\title{GEMS embeddings and freely falling temperatures \\of Schwarzschild(-AdS) black holes in massive gravity }
\author{Soon-Tae Hong}
\email{galaxy.mass@gmail.com}
\affiliation{Center for Quantum Spacetime and
\\ Department of Physics, Sogang University, Seoul 04107, Korea}
\author{Yong-Wan Kim}
\email{ywkim65@gmail.com}
 \affiliation{Department of Physics and
 \\ Research Institute of Physics and Chemistry, Chonbuk National University, Jeonju 54896, Korea}
\author{Young-Jai Park}
\affiliation{Center for Quantum Spacetime and
\\ Department of Physics, Sogang University, Seoul 04107, Korea }
\date{\today}

\begin{abstract}
We globally embed (3+1)-dimensional Schwarzschild and
Schwarzschild-AdS black holes in massive gravity into
(5+2)-dimensional flat spacetimes. Making use of embedding
coordinates, we directly obtain the generalized Hawking, Unruh and
freely falling temperatures in a Schwarzschild and
Schwarzschild-AdS black hole due to massive graviton effects.
\end{abstract}


\maketitle

\section{introduction}
\setcounter{equation}{0}
\renewcommand{\theequation}{\arabic{section}.\arabic{equation}}

It has long been known that any $d$-dimensional
Riemannian manifold can be locally isometrically embedded in an
$N$-dimensional flat one with $N>d$
\cite{Fronsdal:1959zza,Rosen:1965,Goenner1980}. In this respect,
it has been known that the Hawking effect \cite{Hawking:1974sw}
may be related to the Unruh effect \cite{Unruh:1976db}, {\it
i.e.,} the Hawking effect for a fiducial observer in a curved
spacetime can be considered as the Unruh effect for a uniformly
accelerated observer in a higher-dimensional global embedding
Minkowski spacetime (GEMS). Starting from the works of Deser and
Levin \cite{Deser:1997ri,Deser:1998bb,Deser:1998xb}, the GEMS
approach has been used to provide a unified derivation of Unruh
equivalents for Hawking thermal properties
\cite{Hong:2000kn,Kim:2000ct,Hong:2003xz,Chen:2004qw,Santos:2004ws,
Banerjee:2010ma,Cai:2010bv,Hu:2011yx,hong2000,hong2001,
hong2005,hong2004,hong2006,paston2014,Sheykin:2019uwj}. Moreover,
a local temperature measured by a freely falling observer has been
introduced by Brynjolfsson and Thorlacius
\cite{Brynjolfsson:2008uc} using the traditional GEMS method.
Here, a freely falling local temperature is defined at special
turning points of radial geodesics where a freely falling observer
is momentarily at rest with respect to a black hole. After the
work, we have extended the results to other interesting curved
spacetimes \cite{Kim:2009ha,Kim:2013wpa,Kim:2015wwa,hong2015} to
investigate local temperatures of corresponding spacetimes.

On the other hand, Einstein's general relativity (GR) is a
relativistic theory of gravity where the graviton is massless.
However, for many decades, attempts to generalize the Fierz-Pauli
theory \cite{Fierz:1939ix} to a massive gravity theory, which is
reduced to the GR in the massless limit, have been suffered from
difficulty in the presence of the Boulware-Deser ghost
\cite{Boulware:1973my}. Recently, de Rham, Gabadadze, and Trolley
(dRGT) \cite{deRham:2010ik,deRham:2010kj} have obtained a ghost
free massive gravity having a particular type of nonlinear
interaction involving the metric coupled with a symmetric
background tensor, called the reference metric, to create mass
terms. In the dRGT massive gravity, the nondynamical reference
metric is set to be the Minkowskian one, thus breaking the
diffeomorphism invariance which is preserved by GR. It has also
been shown that ghost free massive theories can be obtained using
a general background \cite{Hassan:2011hr,Hassan:2011tf}. For more
details, the readers can refer to the reviews
\cite{Hinterbichler:2011tt,deRham:2014zqa}. Later,  a nonlinear
massive gravity with a special singular reference metric
\cite{Vegh:2013sk} has been studied extensively in the
gauge/gravity duality since the theory breaks the spatial
translational symmetry at the boundary, which is dual to the
gravity theory with broken diffeomorphism invariance in the bulk
\cite{Blake:2013bqa,Amoretti:2014zha,Zhou:2015dha}. Massive
gravity theories with a singular reference metric have also been
exploited to investigate many interesting black hole models
\cite{Cai:2014znn,Adams:2014vza,Hendi:2015pda,Hu:2016hpm,Zou:2016sab,
Hendi:2017fxp,Tannukij:2017jtn,Hendi:2017bys,Hendi:2018xuy,Chabab:2019mlu}.

Very recently, in order to see how massive gravitons affect the
GEMS embeddings and free fall temperatures, we have studied a
charged Ba\~nados-Teitelboim-Zanelli (BTZ) black hole in the
(2+1)-dimensional massive gravity \cite{Hendi:2016pvx}. As a
result, we have explicitly shown that GEMS embedding dimensions
are differently given by a mass parameter \cite{Hong:2018spz}. We
have also obtained the generalized Hawking, Unruh, and freely
falling temperatures of the charged BTZ black hole in massive
gravity theory with massive graviton effects.

In this paper, we will study GEMS embedding of the
Schwarzschild-anti de Sitter (AdS) black hole in a
(3+1)-dimensional massive gravity, and generalize the Hawking,
Unruh, and freely falling temperatures of the Schwarzschild-AdS
black hole in massless gravity to those in massive gravity with
ansatzes in the GEMS approach. In Sec. II, we will study GEMS
embeddings and freely falling temperatures of the Schwarzschild
black hole in massive gravity by comparing it with the
Schwarzschild black hole in massless gravity. In Sec. III, we
present GEMS embeddings of a (3+1)-dimensional Schwarzschild-AdS
black hole in massive gravity into a (5+2)-dimensional flat
spacetime and then newly obtain desired temperatures  $T_{U}$ and
$T_{FFAR}$ of the black holes, which are measured by uniformly
accelerated observers and freely falling ones, respectively.
Finally, conclusions are drawn in Sec. IV.

\section{GEMS of Schwarzschild black hole in massless/massive gravity}
\setcounter{equation}{0}
\renewcommand{\theequation}{\arabic{section}.\arabic{equation}}

\subsection{GEMS of Schwarzschild black hole in massless gravity}

In this section, we briefly recapitulate the GEMS embedding
\cite{Fronsdal:1959zza,Rosen:1965,Goenner1980} and freely falling
temperature \cite{Brynjolfsson:2008uc} of the Schwarzschild black
hole in massless gravity\footnote{In particular, we will call it
massless when $\tilde{m}$ is zero in this work. See the action
(\ref{mSch}) and below.}, which is described by
 \be\label{metric}
 ds^2=-f(r)dt^2+f^{-1}(r)dr^2+r^2(d\theta^2+\sin^2\theta d\phi^2)
 \ee
with the metric function for the massless graviton case
 \be
 f(r)=1-\frac{2m}{r}.
 \ee
From the metric, one can find the surface gravity \cite{wald} as
 \be
 k_H = \left.\sqrt{-\frac{1}{2}(\na^{\mu}\xi^{\nu})(\na_{\mu}\xi_{\nu})}~\right|_{r= r_H}
     =\frac{1}{2r_H},
 \ee
where $\xi^\mu$ is a Killing vector. Then, the Hawking temperature
$T_H$, which is the temperature of the radiation as measured by an
asymptotic observer, is given by
 \be\label{HawkingT}
 T_H=\frac{k_H}{2\pi}=\frac{1}{4\pi r_H}.
 \ee
On the other hand, a local fiducial temperature, measured by an
observer who rests at a distance from a black hole, is given by
 \be\label{fidT}
 T_{\rm FID}(r)=\frac{T_H}{\sqrt{f(r)}}=\frac{r^{1/2}}{4\pi r_H(r-r_H)^{1/2}}.
 \ee
Note that the fiducial temperature $T_{\rm FID}$ diverges at an
event horizon of a black hole, while it becomes the Hawking
temperature asymptotically far away from a black hole.

Now, exploiting the GEMS approach given by the coordinate
transformations for $r\geq r_{H}$, one can embed the
(3+1)-dimensional Schwarzschild spacetime in the massless gravity
(\ref{metric}) into a (5+1)-dimensional flat spacetime as
 \be
 ds^{2}= \eta_{IJ}dz^Idz^J,~{\rm with}~\eta_{IJ}={\rm diag}(-1,1,1,1,1,1).
 \ee
Fronsdal \cite{Fronsdal:1959zza} concretely obtained embedding
functions as follows
 \bea\label{gems-sch}
 z^{0}&=&k_{H}^{-1}f^{1/2}(r)\sinh k_{H}t, \nonumber \\
 z^{1}&=&k_{H}^{-1}f^{1/2}(r)\cosh k_{H}t, \nonumber \\
 z^{2}&=&r\sin\theta\cos\phi, \nonumber \\
 z^{3}&=&r\sin\theta\sin\phi, \nonumber \\
 z^{4}&=&r\cos\theta, \nonumber \\
 z^{5}&=&\int dr
 \left(\frac{r_H(r^2+rr_H+r^2_H)}{r^3}\right)^{1/2}.
 \eea

In static detector ($r,~\theta,~\phi={\rm constrant}$) described
by a fixed point in the ($z^2,~z^3,~z^4,~z^5$) plane on the GEMS
embedded spacetime, an observer, who is uniformly accelerated in
the (5+1)-dimensional flat spacetime, follows a hyperbolic
trajectory in ($z^0,~z^1$) described by
 \be
 a^{-2}_6=(z^1)^2-(z^0)^2= \frac{f(r)}{k^2_H}.
 \ee
Thus, one can arrive at the Unruh temperature for the uniformly
accelerated observer in the (5+1)-dimensional flat spacetime
 \be\label{unruh-sch}
 T_U=\frac{a_6}{2\pi}=\frac{k_H}{2\pi\sqrt{f(r)}}.
 \ee
This corresponds to the fiducial temperature (\ref{fidT}) in the
original Schwarzschild black hole spacetime for an observer
locating at a distance from the black hole. Then, the Hawking
temperature $T_H$ seen by an asymptotic observer can be obtained
as
 \be
 T_H=\sqrt{-g_{00}}T_U=\frac{k_H}{2\pi}.
 \ee
As a result, one can say that the Hawking effect for a fiducial
observer in the black hole spacetime is equal to the Unruh effect
for a uniformly accelerated observer in a higher-dimensional flat
spacetime.

Now, we assume that an observer at rest is freely falling from the
radial position $r=r_{0}$ at
$\tau=0$~\cite{Brynjolfsson:2008uc,Kim:2009ha,Kim:2013wpa,Kim:2015wwa,hong2015}.
The equations of motion for the orbit of the observer are given by
 \bea\label{eomr0}
 \frac{dt}{d\tau}&=&\frac{\sqrt{f(r_0)}}{f(r)},\nonumber\\
 \frac{dr}{d\tau}&=&-[f(r_0)-f(r)]^{1/2}.
 \eea
Exploiting the embedding functions in Eq. (\ref{gems-sch}), one
can easily obtain a freely falling acceleration $\bar{a}_{6}$ in
the (5+1)-dimensional GEMS embedded spacetime as
 \be\label{a6sch}
 \bar{a}^2_{6}=\frac{r^3+r_Hr^2+r^2_Hr+r^3_H}{4r^2_Hr^3},
 \ee
which gives the freely falling temperature at rest (FFAR) measured
by the freely falling observer as
 \be \label{tffar-sch}
 T_{\rm FFAR}=\frac{\bar{a}_6}{2\pi}.
 \ee
Then, by introducing a dimensionless parameter $x=r_H/r$, one can
rewrite a squared freely falling temperature $T^2_{\rm FFAR}$ as
 \be\label{tffar2-sch}
 T^2_{\rm FFAR}=\frac{1+x+x^2+x^3}{16\pi^2r^2_H}.
 \ee
Note that at the event horizon the freely falling temperature
(\ref{tffar-sch}) is finite, while the fiducial temperature
(\ref{unruh-sch}) diverges \cite{Brynjolfsson:2008uc}.

\subsection{GEMS of Schwarzschild black hole in massive gravity}

In this section, we will newly study the GEMS embedding of the
(3+1)-dimensional Schwarzschild black hole in massive gravity,
which is described by the action
 \be\label{mSch}
 S=\frac{1}{16\pi G}\int d^4x\sqrt{-g}\left[{\cal R}
   +{\tilde m}^2\sum_{i=1}^{4}c_i{\cal U}_{i}(g_{\mu\nu},f_{\mu\nu})\right],
 \ee
where ${\cal R}$ is the scalar curvature, $\tilde{m}$ is the
graviton mass, $c_i$ are constants, and ${\cal U}_i$ are symmetric
polynomials of the engenvalue of the matrix ${\cal
K}^\mu_\nu\equiv\sqrt{g^{\mu\alpha}f_{\alpha\nu}}$ as
 \bea
 {\cal U}_1 &=& [{\cal K}],\nonumber\\
 {\cal U}_2 &=& [{\cal K}]^2-[{\cal K}^2],\nonumber\\
 {\cal U}_3 &=& [{\cal K}]^3-3[{\cal K}][{\cal K}^2]+2[{\cal K}^3],\nonumber\\
 {\cal U}_4 &=& [{\cal K}]^4-6[{\cal K}^2][{\cal K}]^2+8[{\cal K}^3][{\cal K}]+3[{\cal K}^2]^2-6[{\cal K}^4].
 \eea
Here, the square root in ${\cal K}$ means
$(\sqrt{A})^\mu_\alpha(\sqrt{A})^\alpha_\nu=A^\mu_\nu$ and $[{\cal
K}]$ denotes the trace ${\cal K}^\mu_\mu$. Finally, $f_{\mu\nu}$
is a non-dynamical, fixed symmetric tensor, called the reference
metric, introduced to construct nontrivial interaction terms in
massive gravity.

Variation of the action with respect to $g^{\mu\nu}$ gives us the
equations of motion as
 \be\label{eom}
 {\cal R}_{\mu\nu}-\frac{1}{2}{\cal R}g_{\mu\nu}+\tilde{m}^2\chi_{\mu\nu}=0,
 \ee
where
 \bea
 \chi_{\mu\nu}&=&-\frac{c_1}{2}({\cal U}_1g_{\mu\nu}-{\cal K}_{\mu\nu})
    -\frac{c_2}{2}({\cal U}_2g_{\mu\nu}-2{\cal U}_1{\cal K}_{\mu\nu}+2{\cal K}^2_{\mu\nu})
    -\frac{c_3}{2}({\cal U}_3g_{\mu\nu}-3{\cal U}_2{\cal K}_{\mu\nu}+6{\cal U}_1{\cal K}^2_{\mu\nu}
       -6{\cal K}^3_{\mu\nu}) \nonumber\\
    &-&\frac{c_4}{2}({\cal U}_4g_{\mu\nu}-4{\cal U}_3{\cal K}_{\mu\nu}+12{\cal U}_2{\cal K}^2_{\mu\nu}
       -24{\cal U}_1{\cal K}^3_{\mu\nu}+24{\cal K}^4_{\mu\nu}).
 \eea
We consider the spherically symmetric black hole solution ansatz
of \be\label{metric-mSch}
 ds^2=-f(r)dt^2+f^{-1}(r)dr^2+r^2(d\theta^2+\sin^2\theta d\phi^2),
 \ee
and the special form of the reference metric as
 \be\label{fidmetric}
 f_{\mu\nu}={\rm diag}(0,0,c^2_0,c^2_0\sin^2\theta),
 \ee
where $c_0$ is a positive constant
\cite{Vegh:2013sk,Blake:2013bqa,Amoretti:2014zha,Zhou:2015dha,Cai:2014znn,Adams:2014vza}.

It seems appropriate to comment on the action having this
particular reference metric form. Although such a theory breaks
global Lorentz invariance and Copernican principle, this action is
still invariant under diffeomorphism in the ($t,r$) plane, but not
in the ($\theta,\phi$) plane. This implies at the dual side that
the theory has conserved energy but no conserved momentum
\cite{Blake:2013bqa,Amoretti:2014zha,Zhou:2015dha,Cai:2014znn,Adams:2014vza}.

Now, one can find symmetric potentials as
 \be
 {\cal U}_1=\frac{2c_0}{r},~~~{\cal U}_2=\frac{2c^2_0}{r^2},~~~{\cal U}_3={\cal
 U}_4=0.
 \ee
Then, the $tt$($rr$)- and $\theta\theta$($\phi\phi$)-components of
the equations of motion (\ref{eom}) are reduced to
 \bea
 f'(r)+\frac{f(r)}{r}-c_0c_1\tilde{m}^2-\frac{1+c^2_0c_2\tilde{m}^2}{r}&=&0,\\
 f''(r)+\frac{2}{r}f'(r)-\frac{c_0c_1\tilde{m}^2}{r}&=&0,
 \eea
respectively. Here we note that the
$\theta\theta$($\phi\phi$)-components can be obtained by
differentiating the $tt$($rr$)-components with respect to $r$. As
a result, we have the solution as follows
 \be
 f(r)=1-\frac{2m}{r}+\frac{c_0c_1\tilde{m}^2}{2}r+c^2_0c_2\tilde{m}^2,
 \ee
where $m$ is an integration constant related to the mass of the
black hole. Without loss of generality, one can define
$R=c_0c_1\tilde{m}^2/4$ and ${\cal C}=c^2_0c_2\tilde{m}^2$, and
finally obtain the solution of
 \be\label{lapsewol}
 f(r)=1-\frac{2m}{r}+2Rr+{\cal C}.
 \ee

Now, from the solution (\ref{lapsewol}), one can find the surface
gravity~\cite{wald}
 \be\label{kh-msch}
 k_H =\frac{1+{\cal C}}{2r_{H}}+2R,
 \ee
the Hawking temperature $T_H$
 \be\label{HTmSch}
 T_H=\frac{1+{\cal C}}{4\pi r_H}+\frac{R}{\pi},
 \ee
and the local fiducial temperature $T_{\rm FID}$
 \be\label{fidTmSch}
 T_{\rm FID}=\frac{T_H}{\sqrt{f(r)}}=\frac{r^{1/2}(1+{\cal C}+4Rr_H)}
 {4\pi r_H(r-r_H)^{1/2}[1+{\cal C}+2R(r+r_H)]^{1/2}}.
 \ee
According to the values of ${\cal C}$ and $R$, the Hawking
temperature (\ref{HTmSch}) in the massive gravity differently
behaves, which is shown in Fig. 1. Note that first $R$ gives a
constant contribution to the Hawking temperature. The Hawking
temperature is mostly proportional to $1/r_H$. When ${\cal C}>-1$,
it decreases as $r_H$ increases. When ${\cal C}=-1$, it is just a
constant given by $R$. However, when ${\cal C}<-1$, it is
negatively proportional to $1/r_H$. Since the Hawking temperature
is flipped according to the sign of $\cal C$, it should be careful
to embed this spacetime into a flat one by using the GEMS
approach, which will be now discussed in the following.

\begin{figure*}[t!]
   \centering
   \includegraphics{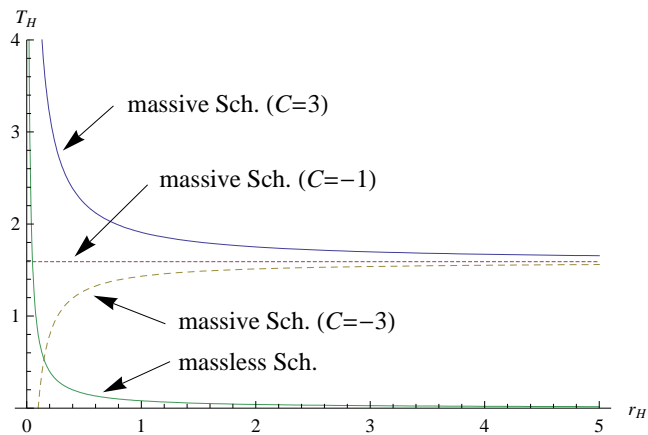}
\caption{Hawking temperature for Schwarzschild black hole in
massive gravity with varying $\cal C$ and $R=5$, comparing with
one in massless gravity.}
 \label{fig1}
\end{figure*}

Now, let us newly exploit the GEMS approach given by the
coordinate transformations for $r\geq r_{H}$ and $R\ge 0$ as in
Sec. II.A. As a result, we have found that the (3+1)-dimensional
Schwarzschild spacetime in the massive gravity (\ref{metric-mSch})
can be embedded into a (5+2)-dimensional flat one, depending on
the massive parameter $\cal C$ as
 \be
 ds^{2}= \eta_{IJ}dz^Idz^J,~{\rm with}~\eta_{IJ}={\rm diag}(-1,1,1,1,1,1,-1),
 \ee
where the embedding functions are found to be
 \bea \label{gemsmsch}
 z^{0}&=&k_{H}^{-1}f^{1/2}(r)\sinh k_{H}t, \nonumber \\
 z^{1}&=&k_{H}^{-1}f^{1/2}(r)\cosh k_{H}t, \nonumber \\
 z^{2}&=&r\sin\theta\cos\phi, \nonumber \\
 z^{3}&=&r\sin\theta\sin\phi, \nonumber \\
 z^{4}&=&r\cos\theta
 \eea
with
 \bea\label{posC}
 z^5 &=& \int dr
         \left(\frac{r_H(1+h_1+{\cal C}h_2)h_5}
             {r^3h_{6}h^2_7}\right)^{1/2},
         \nonumber\\
 z^6&=&\int dr
    \left(\frac{2R(r+r_H)(1+h_1+{\cal C}h_2)+4r^2_H(h_3+{\cal C}h_4)}
               {h_{6}h^2_7}\right)^{1/2},
 \eea
when ${\cal C}>0$, and
 \bea\label{negC}
 z^5 &=& \int dr
         \left(\frac{r_H(1+h_1)h_5}{r^3h_{6}h^2_7}
              -\frac{2{\cal C}R(r+r_H)h_2+4{\cal C}r^2_Hh_4}{h_{6}h^2_7}
         \right)^{1/2},
         \nonumber\\
 z^6&=&\int dr
   \left(\frac{2R(r+r_H)(1+h_1)+4r^2_Hh_3}{h_{6}h^2_7}-\frac{{\cal C}r_Hh_2h_5}{r^3h_{6}h^2_7}
         \right)^{1/2},
 \eea
when ${\cal C}<0$. Here, we have defined some functions associated
with $z^5$ and $z^6$ as follows
 \bea\label{someftns}
 h_1&=&8Rr_H(1+2Rr_H)+{\cal C}^2, \nonumber\\
 h_2&=&2(1+4Rr_H), \nonumber\\
 h_3&=&\frac{R^2(r^2+3r^2_H)(r+r_H)}{r^3}, \nonumber\\
 h_4&=&\frac{Rr_H(r+r_H)}{r^3}+\frac{(1+4Rr_H)^2}{4r^2_H}
                +\frac{{\cal C}^2}{4r^2_H},\nonumber\\
 h_5&=&r^2+r_Hr+r^2_H,\nonumber\\
 h_{6}&=&1+{\cal C}+2R(r+r_H),\nonumber\\
 h_7&=&1+{\cal C}+4Rr_H.
 \eea
Moreover, we have imposed the restriction that the function $h_6$
is positive definite. Note that in the limit of both ${\cal
C}\rightarrow 0$ and $R\rightarrow 0$, the timelike embedding
coordinates $z^6$ in Eqs. (\ref{posC}) and (\ref{negC}) become
zero, regardless of the sign of $\cal C$. As a result, the
(5+2)-dimensional flat spacetimes in the Schwarzschild black hole
in the massive gravity are reduced to the (5+1)-dimensional flat
ones in the massless Schwarzschild black hole exactly
\cite{Deser:1998xb,Kim:2000ct,Brynjolfsson:2008uc,Kim:2009ha}.

Now, in static detectors ($r$, $\theta$, $\phi=$const) described
by a fixed point in the ($z^{2}$, $z^{3}$, $z^{4}$, $z^{5}$,
$z^{6}$) plane, a uniformly accelerated observer in the
(5+2)-dimensional flat spacetime, follows a hyperbolic trajectory
in ($z^{0}$,$z^{1}$) described by a proper acceleration $a_{7}$ as
follows
 \be\label{acc-msch}
 a^{-2}_{7}=(z^1)^2-(z^0)^2
  =\frac{4r^2_H(r-r_H)[1+{\cal C}+2R(r+r_H)]}
              {r(1+{\cal C}+4Rr_H)^2}.
 \ee
Thus, we arrive at the Unruh temperature for the uniformly
accelerated observer in the (5+2)-dimensional flat spacetime
 \be\label{ut-msch}
 T_U=\frac{a_7}{2\pi}=\frac{r^{1/2}(1+{\cal C}+4Rr_H)}
 {4\pi r_H(r-r_H)^{1/2}[1+{\cal C}+2R(r+r_H)]^{1/2}}.
 \ee
This is exactly the same with the local fiducial temperature
(\ref{fidTmSch}) measured by a local observer rest at a distance
from the black hole.

Now, we assume that an observer at rest is freely falling from the
radial position $r=r_{0}$ at
$\tau=0$~\cite{Brynjolfsson:2008uc,Kim:2009ha,Kim:2013wpa,Kim:2015wwa,hong2015}.
The equations of motion for the orbit of the observer are given by
Eq. (\ref{eomr0}). Exploiting Eqs. (\ref{gemsmsch}), (\ref{posC})
for ${\cal C}>0$ (or, (\ref{negC}) for ${\cal C}<0$) with Eq.
(\ref{eomr0}), we obtain a freely falling acceleration
$\bar{a}_{7}$ in the (5+2)-dimensional GEMS embedded spacetime for
the Schwarzschild black hole in the massive gravity as
 \be\label{a7msch}
 \bar{a}^2_{7}=\frac{(r+r_H)(1+{\cal C}+2Rr_H)[(r^2+r^2_H)(1+{\cal C}+2Rr_H)+4Rr_Hr^2]}
              {4r^2_Hr^3[1+{\cal C}+2R(r+r_H)]},
 \ee
which gives us a freely falling local temperature measured by the
freely falling observer as
 \be
 T_{\rm FFAR}=\frac{\bar{a}_7}{2\pi}.
 \label{tffar}
 \ee
By using the dimensionless parameters $x=r_H/r$ and $d=Rr_H$, the
squared freely falling temperature $T^2_{\rm FFAR}$ becomes
 \be\label{tffarmsch}
 T^2_{\rm FFAR}=\frac{(1+{\cal C}+2d)[(1+{\cal C}+2d)(1+x+x^2+x^3)+4d(1+x)]x}
                    {16\pi^2r^2_H[(1+{\cal C}+2d)x+2d]},
 \ee
which remains finite at the event horizon. In Fig. 2, we have
depicted the ratio of the squared freely falling temperature
$T^2_{\rm FFAR}/T^2_H$ for the Schwarzschild black hole in the
massless/massive gravity in the range of $0<x\le 1$. Note that at
the event horizon of $x=1$ ($r=r_H$) the freely falling
temperatures are all finite.

\begin{figure*}[t!]
   \centering
   \includegraphics{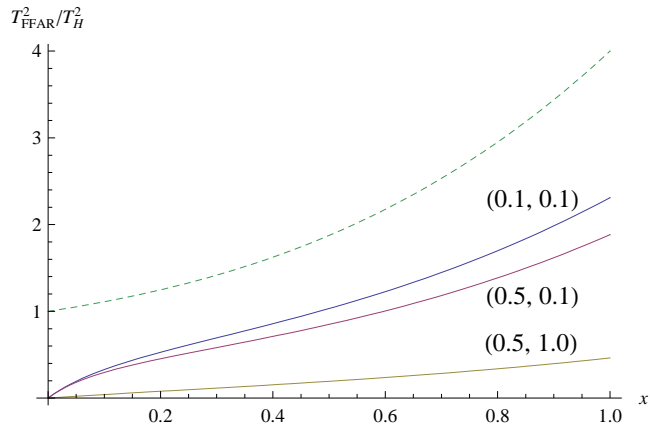}
\caption{Squared freely falling temperatures for Schwarzschild
black hole in massive gravity for $({\cal C}, d)=(0.1,0.1),~
(0.5,0.1),~(0.5,1.0)$. The dashed line is for Schwarzschild black
hole in massless gravity.}
 \label{fig2}
\end{figure*}

\section{GEMS of Schwarzschild-AdS black hole in massless/massive gravity}
\setcounter{equation}{0}
\renewcommand{\theequation}{\arabic{section}.\arabic{equation}}

\subsection{GEMS of Schwarzschild-AdS black hole in massless gravity}

In this section, we also briefly summarize the GEMS embeddings
\cite{Fronsdal:1959zza,Rosen:1965,Goenner1980} and freely falling
temperature \cite{Brynjolfsson:2008uc} of the Schwarzschild-AdS
black hole in massless gravity, which is described by
 \be\label{metric-schads}
 ds^2=-f(r)dt^2+f^{-1}(r)dr^2+r^2(d\theta^2+\sin^2\theta d\phi^2)
 \ee
with the metric function for the massless graviton case
 \be
 f(r)=1-\frac{2m}{r}+\frac{r^2}{l^2}.
 \ee
This Schwarzschild-AdS solution has the surface gravity
 \be
 k_H =\frac{1}{2r_{H}}+\frac{3r_H}{2l^2},
 \ee
the Hawking temperature $T_H$
 \be
 T_H=\frac{1}{4\pi r_{H}}+\frac{3r_{H}}{4\pi l^{2}},
 \ee
and the local fiducial temperature $T_{\rm FID}$ as
 \be\label{fidTsads}
 T_{\rm FID}=\frac{r^{1/2}(3r^2_H+l^2)}{4\pi r_Hl(r-r_H)^{1/2}(r^2+rr_H+r^2_H+l^2)^{1/2}}.
 \ee

Exploiting the GEMS approach given by the coordinate
transformations for $r\geq r_{H}$, we can embed the
(3+1)-dimensional Schwarzschild-AdS black hole in the massless
gravity (\ref{metric-schads}) into a (5+2)-dimensional flat
spacetime as
 \be
 ds^{2}= \eta_{IJ}dz^Idz^J,~{\rm with}~\eta_{IJ}={\rm diag}(-1,1,1,1,1,1,-1),
 \ee
where the embedding functions are given by
 \bea\label{gems-schads}
 z^{0}&=&k_{H}^{-1}f^{1/2}(r)\sinh k_{H}t, \nonumber \\
 z^{1}&=&k_{H}^{-1}f^{1/2}(r)\cosh k_{H}t, \nonumber \\
 z^{2}&=&r\sin\theta\cos\phi, \nonumber \\
 z^{3}&=&r\sin\theta\sin\phi, \nonumber \\
 z^{4}&=&r\cos\theta, \nonumber \\
 z^{5}&=&\int dr \frac{l(r^2_H+l^2)}{3r^2_H+l^{2}}\left(\frac{r_Hh_5}{r^3(h_5+l^2)}\right)^{1/2},\nonumber\\
 z^{6}&=&\int dr \frac{1}{3r^2_H+l^2}\left(\frac{h_5 h_8}{h_5+l^2}\right)^{1/2},
 \eea
where $h_5$ is given by Eq. (\ref{someftns}) and $h_8$ is defined
as
 \be\label{someftn1}
 h_8=9r^4_H+10r^2_Hl^2+l^4.
 \ee

In static detector ($r,~\theta,~\phi={\rm constrant}$) described
by a fixed point in the ($z^2,~z^3,~z^4,~z^5,~z^6$) plane, a
uniformly accelerated observer in the (5+2)-dimensional flat
spacetime, follows a hyperbolic trajectory in ($z^0,~z^1$)
described by
 \be
 a^{-2}_7=(z^1)^2-(z^0)^2= \frac{4r^2_Hl^2(r-r_H)(r^2+rr_H+r^2_H+l^2)}{r(3r^2_H+l^2)^2}.
 \ee
Thus, we arrive at the Unruh temperature for a uniformly
accelerated observer in the (5+2)-dimensional flat spacetime
 \be\label{ut-schads}
 T_U=\frac{a_7}{2\pi}=\frac{r^{1/2}(3r^2_H+l^2)}{4\pi r_Hl(r-r_H)^{1/2}(r^2+rr_H+r^2_H+l^2)^{1/2}}.
 \ee
This corresponds to the local fiducial temperature
(\ref{fidTsads}) in the original Schwarzschild-AdS black hole
spacetime for an observer locating at a distance from the black
hole, and one can find the Hawking temperature $T_H$ seen by an
asymptotic observer as
 \be
 T_H=\sqrt{-g_{00}}T_U=\frac{k_H}{2\pi}.
 \ee
As before, one can see that the Hawking effect for a fiducial
observer in the black hole spacetime is equal to the Unruh effect
for a uniformly accelerated observer in a higher-dimensional flat
spacetime.

Now, we assume that an observer at rest is freely falling from the
radial position $r=r_{0}$ at
$\tau=0$~\cite{Brynjolfsson:2008uc,Kim:2009ha,Kim:2013wpa,Kim:2015wwa,hong2015}.
The equations of motion for the orbit of the observer are given by
Eq. (\ref{eomr0}), and by exploiting Eq. (\ref{gems-schads}), we
obtain a freely falling acceleration $\bar{a}_{7}$ in the
(5+2)-dimensional GEMS embedded spacetime as
 \be\label{a2schads}
  \bar{a}^2_{7}=\frac{[(r+r_H)(r^2_H+l^2)-2r_Hr^2][(r^2+r^2_H)(r^2_H+l^2)+2r_Hr^2(r+r_H)]}
              {4r^2_Hr^3l^2(r^2+r_Hr+r^2_H+l^2)}.
 \ee
This gives us a freely falling temperature measured by the freely
falling observer as
 \be \label{tffar-schads}
 T_{\rm FFAR}=\frac{\bar{a}_7}{2\pi}.
 \ee
By using the dimensionless parameter $x=r_H/r$, $c=l/r_H$, the
squared freely falling temperature $T^2_{\rm FFAR}$ can be written
as
 \be\label{ffar2schads}
 T^2_{\rm FFAR}=\frac{-4(1+x)+(c^2+1)(c^2+5)x^2+(c^2+1)^2(1+x+x^2)x^3}{16\pi^2l^2[1+x+(c^{2}+1)x^{2}]}.
 \ee
In Fig. 3, we have depicted squared freely falling temperatures
for the Schwarzschild-AdS black hole in the massless gravity.
\begin{figure*}[t!]
   \centering
   \includegraphics{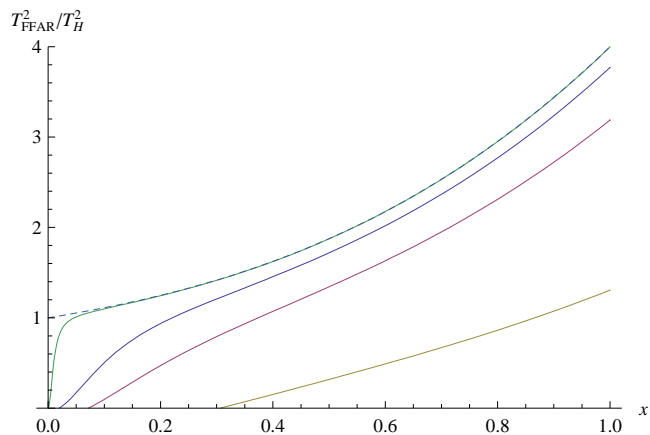}
\caption{Squared freely falling temperatures for the
Schwarzschild-AdS black hole in masless gravity with
$c=2,~5,~10,~100$ from bottom to top. Here, the dashed line is for
the Schwarzschild black hole in massless gravity.}
 \label{fig3}
\end{figure*}
Note that in the limit of $x\rightarrow 1$ (or, $r\rightarrow
r_H$), the freely falling temperatures $T_{\rm FFAR}$ are finite,
while the Unruh temperature (\ref{ut-schads}) diverges. Moreover,
at the spatial infinity limit of $x\rightarrow 0$, one can obtain
imaginary temperature as
 \be
  T^2_{\rm FFAR}=\frac{-1}{4\pi^2l^2}<0,
 \ee
which is seemingly unphysical. However, it is allowed for a
geodesic observer who follows a spacelike motion
\cite{Deser:1997ri,Brynjolfsson:2008uc,Kim:2009ha}.

\subsection{GEMS of Schwarzschild-AdS black hole in massive gravity}

In this section, we will finally study the following
(3+1)-dimensional massive gravity with a negative cosmological
constant
as~\cite{Vegh:2013sk,Cai:2014znn,Hendi:2015pda,Hendi:2016pvx}
 \be\label{mSchads}
 S=\frac{1}{16\pi G}\int d^4x\sqrt{-g}\left[{\cal R}
   -\Lambda +{\tilde m}^2\sum_{i=1}^{4}c_i{\cal
   U}_{i}(g_{\mu\nu},f_{\mu\nu})\right],
 \ee
where $\Lambda=-6/l^2$ and others are the same as before. As in
Sec.II.B, from the same ansatz of the reference metric
$f_{\mu\nu}={\rm diag}(0,0,c^2_0,c^2_0\sin^2\theta)$ and the
spherically symmetric metric ansatz of
 \be\label{metric-mschads}
 ds^2=-f(r)dt^2+f^{-1}(r)dr^2+r^2(d\theta^2+\sin^2\theta d\phi^2),
 \ee
one has the $tt(rr)$- and $\theta\theta$($\phi\phi$)-components of
the equations of motion as follows
 \bea\label{tteom}
 f'(r)+\frac{f(r)}{r}-\frac{3r}{l^2}-c_0c_1\tilde{m}^2-\frac{1+c^2_0c_2\tilde{m}^2}{r}&=&0,\\
 \label{thetathetaeom}
 f''(r)+\frac{2}{r}f'(r)-\frac{c_0c_1\tilde{m}^2}{r}-\frac{6}{l^2}&=&0,
 \eea
respectively. These equations include the cosmological constant
contribution. Note also that (\ref{thetathetaeom}) can be obtained
by differentiating (\ref{tteom}) with respect to $r$. As a result,
the solution can be found as
 \be
 f(r)=1-\frac{2m}{r}+\frac{r^2}{l^2}+\frac{c_0c_1\tilde{m}^2}{2}r+c^2_0c_2\tilde{m}^2.
 \ee
As before, one can define $R=c_0c_1\tilde{m}^2/4$ and ${\cal
C}=c^2_0c_2\tilde{m}^2$, and finally obtain the solution as
follows
 \be\label{lapse}
 f(r)=1-\frac{2m}{r}+\frac{r^2}{l^2}+2Rr+{\cal C}.
 \ee
This spacetime is asymptotically described by the AdS. However,
the vacuum solution with $m=0$ is not an AdS unless ${\tilde m}=0$
($ \it{i.e.}~R={\cal C}$=0)  in Eq. (\ref{mSchads}). Since the
event horizon is determined by $f(r)|_{r=r_H}=0$, the mass $m$ can
be written in terms of the event horizon $r_{H}$ as
 \be\label{mass}
  m(r_H)=\frac{(1+{\cal C})r_H}{2}+Rr^2_H+\frac{r_{H}^{3}}{2l^{2}}.
 \ee
In Fig. 4, we have drawn the metric $f(r)$ and mass functions
$m(r_{H})$ to compare the features of the Schwarzschild-AdS black
hole in the massive gravity with those in the massless gravity.

\begin{figure*}[t!]
   \centering
   \includegraphics{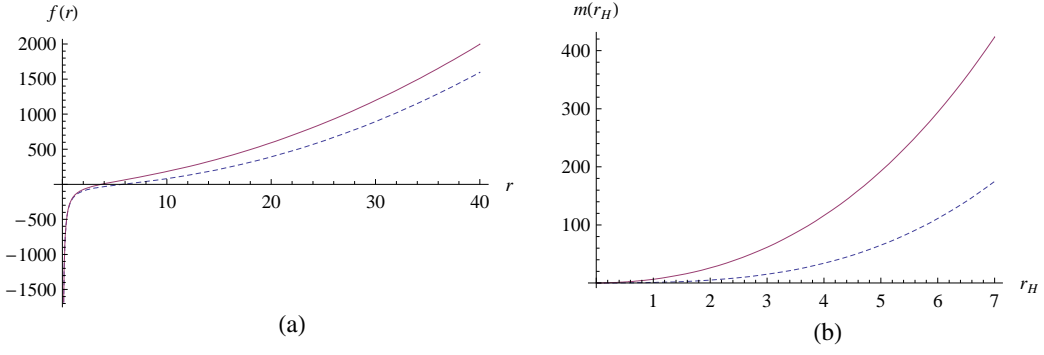}
\caption{For the Schwarzschild-AdS black hole in massive gravity,
(a) metric function with $m=100$, $l=1$, $R=5$, and ${\cal C}=1$,
and (b) mass function $m(r_H)$ with $l=1$, $R=5$, and ${\cal
C}=1$. For comparison, metric and mass functions for the
Schwarzschild-AdS black hole in massless gravity are drawn by
dashed lines with $l=1$, $R=0$, and ${\cal C}=0$.}
 \label{fig4}
\end{figure*}

\begin{figure*}[t!]
 \centering
 \includegraphics{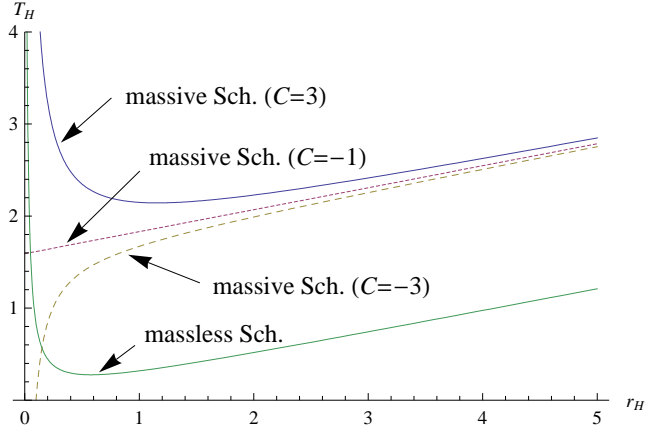}
 \caption{Hawking temperatures for the Schwarzschild-AdS black hole in
 massive gravity with varying $\cal C$ and $l=1$, $R=5$, comparing with
 the one in massless gravity.}
 \label{fig5}
\end{figure*}

From the solution, one can find the surface gravity as
 \be\label{kh0}
 k_H =\frac{1+{\cal C}}{2r_H}+2R+\frac{3r_H}{2 l^2} ,
 \ee
the Hawking temperature $T_H$
 \be\label{HT-mschads}
 T_H=\frac{1+{\cal C}}{4\pi r_H}+\frac{R}{\pi}+\frac{3r_H}{4\pi l^2},
 \ee
and a local fiducial temperature $T_{\rm FID}$ as
 \be\label{fidT-mschads}
 T_{\rm FID}=\frac{r^{1/2}[3r^2_H+(1+{\cal C}+4Rr_H)l^2]}
 {4\pi r_Hl(r-r_H)^{1/2}[r^2+r_Hr+r^2_H+(1+{\cal C})l^2+2R(r+r_H)l^2]^{1/2}}.
 \ee

In Fig. 5, we have drawn the Hawking temperatures for the
Schwarzschild-AdS black hole in the massive gravity, compared with
the case of the massless gravity. For the case of ${\cal C}>-1$,
the Hawking temperature in the massive gravity has a minimum at
$r_H=r_m=l\sqrt{(1+{\cal C})/3}$. When ${\cal C}=-1$, the Hawking
temperature is a straight line having a minimum value $R/\pi$ at
$r_H=0$. When ${\cal C}<-1$, since the first term in Eq.
(\ref{HawkingT}) has a minus sign, the Hawking temperature
monotonically decreases as $r_H$ decreases, and finally becomes
zero at $r_H=r_{0}=\frac{2}{3}Rl^2\left(\sqrt{1-\frac{3(1+{\cal
C})}{4R^2l^2}}-1\right)$. In short, the Hawking temperature in the
massive gravity can be classified by the values of ${\cal C}$ and
$R$: when ${\cal C}>-1$, the Hawking temperature behaves as the
Schwarzschild-AdS black hole in the massless gravity. When ${\cal
C}=-1$, the Hawking temperature is a straight line, and when
${\cal C}<-1$, the Hawking temperature becomes a monotonically
decreasing function as $r_H$ decreases. Additionally, $R$ just
shifts the Hawking temperatures vertically up and down.

Now, exploiting the GEMS approach given by the coordinate
transformations for $r\geq r_{H}$ and $R\ge 0$, we can embed the
(3+1)-dimensional Schwarzschild-AdS in the massive gravity
(\ref{metric-mschads}) into a (5+2)-dimensional flat spacetime as
 \be
 ds^{2}= \eta_{IJ}dz^Idz^J,~{\rm with}~\eta_{IJ}={\rm diag}(-1,1,1,1,1,1,-1),
 \ee
where embedding functions are found to be
 \bea\label{gemsmschads0}
 z^{0}&=&k_{H}^{-1}f^{1/2}(r)\sinh k_{H}t, \nonumber \\
 z^{1}&=&k_{H}^{-1}f^{1/2}(r)\cosh k_{H}t, \nonumber \\
 z^{2}&=&r\sin\theta\cos\phi, \nonumber \\
 z^{3}&=&r\sin\theta\sin\phi, \nonumber \\
 z^{4}&=&r\cos\theta
 \eea
with
 \bea\label{gemsmschads1}
 z^5 &=& l\int dr
         \left(\frac{((r^2_H+l^2)^2+l^4(h_1+{\cal C}h_2))r_Hh_5}
             {r^3(h_5+l^2 h_{6})h^2_9}\right)^{1/2},   \nonumber\\
 z^6&=&\int dr
    \left(\frac{(h_7+l^4(h_1+{\cal C}h_2))(h_5+2Rl^2(r+r_H))+4l^6r^2_H(h_3+{\cal C}h_4)}
               {(h_5+l^2h_{6})h^2_9}\right)^{1/2},
 \eea
when ${\cal C}>0$, and
 \bea\label{gemsmschads2}
 z^5 &=& l\int dr
         \left(\frac{l^2((r^2_H+l^2)^2+l^4h_1)r_Hh_5}{r^3(h_5+l^2h_{6})h^2_9}
             -\frac{{\cal C}l^4h_2(h_5+2Rl^2(r+r_H)) +4{\cal C}l^6r^2_Hh_4}{(h_5+l^2 h_{6})h^2_9}\right)^{1/2},
         \nonumber\\
 z^6&=&\int dr
    \left(\frac{(h_7+l^4h_1)(h_5+2Rl^2(r+r_H))+4l^6r^2_Hh_3}{(h_5+l^2h_{6})h^2_9}-\frac{{\cal C}l^4r_Hh_2h_5}{r^3(h_5+l^2h_{6})h^2_9}\right)^{1/2},
 \eea
when ${\cal C}<0$. Associated with $z^5$ and $z^6$ coordinates in
Eqs. (\ref{gemsmschads1}) and (\ref{gemsmschads2}), we have used the
functions, $h_{5}$, $h_6$ and $h_7$ given in Eqs. (\ref{someftns}) and (\ref{someftn1}),
and newly defined functions as
 \bea
 h_1&=&8Rr_H\left(1+2Rr_H+\frac{3r^2_H}{l^2}\right)+{\cal C}^2, \nonumber\\
 h_2&=&2\left(1+4Rr_H+\frac{3r^2_H}{l^2}\right), \nonumber\\
 h_3&=&\frac{4Rr^2_H}{l^2r}+\frac{Rr_H(5r^2_H+l^2)(r+r_H)}{l^2r^3}+\frac{R^2(r^2+3r^2_H)(r+r_H)}{r^3}
      +\frac{{\cal C}^2(3r^2_H+(1+4Rr_H)l^2)}{2r^2_Hl^2},\nonumber\\
 h_4&=&\frac{r_H(r^2+r_Hr+r^2_H)}{l^2r^3}+\frac{Rr_H(r+r_H)}{r^3}
     +\frac{(3r^2_H+(1+4Rr_H)l^2)^2}{4r^2_Hl^4}+\frac{{\cal C}^2}{4r^2_H},\nonumber\\
 h_9&=&3r^2_H+(1+{\cal C}+4Rr_H)l^2.
 \label{funcs}
 \eea
Here, we have imposed the restriction that the function
$h_5+l^2h_{6}$ is positive definite.

In static detectors ($\theta$, $\phi$, $r=$ const) described by a
fixed point in the ($z^{2}$, $z^{3}$, $z^{4}$, $z^{5}$, $z^{6}$)
plane, a uniformly accelerated observer in the (5+2)-dimensional
flat spacetime follows a hyperbolic trajectory in
($z^{0}$,$z^{1}$) described by a proper acceleration $a_{7}$ as
follows
 \be\label{acc2-mschads}
 a^{-2}_{7}=(z^1)^2-(z^0)^2
  =\frac{4r^2_Hl^2(r-r_H)[r^2+r_Hr+r^2_H+(1+{\cal C})l^2+2R(r+r_H)l^2]}
              {r[3r^2_H+(1+{\cal C}+4Rr_H)l^2]^2}.
 \ee
Thus, we arrive at the Unruh temperature for a uniformly
accelerated observer in the (5+2)-dimensional flat spacetime as
 \be\label{ut-mschads}
 T_U=\frac{a_7}{2\pi}=\frac{r^{1/2}[3r^2_H+(1+{\cal C}+4Rr_H)l^2]}
 {4\pi r_Hl(r-r_H)^{1/2}[r^2+r_Hr+r^2_H+(1+{\cal C})l^2+2R(r+r_H)l^2]^{1/2}}.
 \ee
This is exactly the same with the fiducial temperature
(\ref{fidT-mschads}), showing the fact that the Hawking effect for
a fiducial observer in the black hole spacetime is equal to the
Unruh effect for a uniformly accelerated observer in a
higher-dimensional flat spacetime.

Now, we assume that an observer at rest is freely falling from the
radial position $r=r_{0}$ at
$\tau=0$~\cite{Brynjolfsson:2008uc,Kim:2009ha,Kim:2013wpa,Kim:2015wwa,hong2015}.
The equations of motion for the orbit of the observer are given by
Eq. (\ref{eomr0}). Exploiting Eqs. (\ref{gemsmschads0}),
(\ref{gemsmschads1}) for ${\cal C}>0$ (or, (\ref{gemsmschads2})
for ${\cal C}<0$) and (\ref{eomr0}), we obtain a freely falling
acceleration $\bar{a}_{7}$ in the (5+2)-dimensional GEMS embedded
spacetime
 \be\label{a7mschads}
 \bar{a}^2_{7}=\frac{[(r+r_H)(r^2_H+l^2(1+{\cal C}+2Rr_H))-2r_Hr^2][(r^2+r^2_H)(r^2_H+l^2(1+{\cal C}+2Rr_H))+2r_Hr^2(r+r_H)+4Rr_Hr^2l^2]}
              {4r^2_Hr^3l^2[r^2+r_Hr+r^2_H+l^2(1+{\cal C}+2R(r+r_H))]},
 \ee
which gives us the temperature measured by the freely falling
observer as
 \be \label{tffar-mschads}
 T_{\rm FFAR}=\frac{\bar{a}_7}{2\pi}.
 \ee
With previously defined massless parameters of $x=r_H/r$,
$c=l/r_H$, and $d=Rr_H$, the squared freely falling temperature
$T^2_{\rm FFAR}$ can be rewritten as
 \be\label{tffar2mschads}
 T^2_{\rm FFAR}=\frac{g(x,c,{\cal C},d)}{16\pi^2r^2_Hc^2[1+x+(1+c^2)x^2+2c^2d(1+x)x+{\cal C}c^2x^2]},
 \ee
where
 \bea
  g(x,c,{\cal C},d)&=&-4(1+x+2c^2dx)+(1+c^2+2c^2d+{\cal C}c^2)(5+c^2+2c^2d+{\cal C}c^2)x^2 \nonumber\\
                 && +(1+c^2+2c^2d+{\cal C}c^2)^2(1+x+x^2)x^3+4c^2d(1+c^2+2c^2d+{\cal
                 C}c^2)(1+x)x^2.
 \eea
The ratio of the squared freely falling temperature for the
Schwarzschild-AdS black hole in the massive gravity to the squared
Hawking temperature is plotted in Fig. 6. Note that the freely
falling temperatures are finite at the event horizons. In order to
see the graviton mass effect clearly, we have redrawn Fig. 6 to
Fig. 7 by rescaling ten times $T^2_{\rm FFAR}/T^2_H$ with $c=1$.
Then, in Fig. 7(a), one can see that the freely falling
temperature starts to appear when ${\cal C}>-1$. One can also see
that the graviton mass effect of $({\cal C}, d)$ on the freely
falling temperatures is increased as the freely falling observer
approaches to the event horizon. Moreover, the squared freely
falling temperatures near the event horizon are more enhanced when
$({\cal C}, d)$ are relatively small.

\begin{figure*}[t!]
 \centering
 \includegraphics{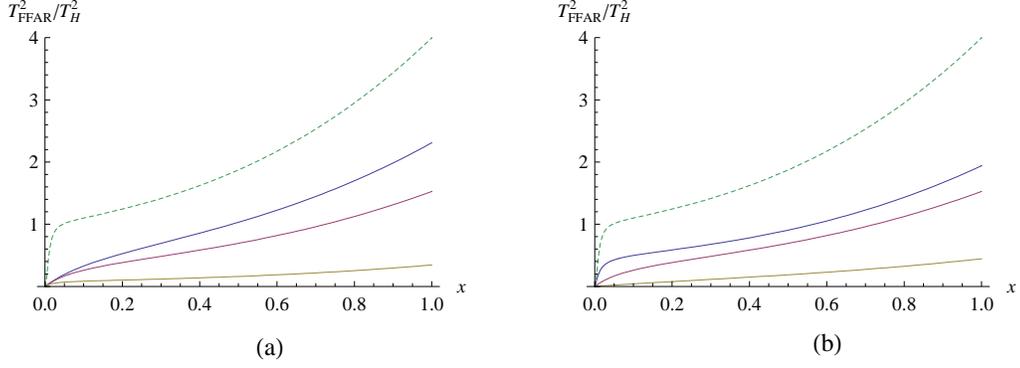}
  \caption{Freely falling temperatures for the Schwarzschild-AdS black hole in massive gravity
 (a) with varying ${\cal C}=0.1,~1,~10$ from top to bottom for fixed $c=100$, $d=0.1$,
 and (b) with varying $d=0.01,~0.1,~1$  from top to bottom for fixed ${\cal C}=1$, $c=100$.
 Here, the dotted lines are for the Schwarzschild-AdS black hole in massless gravity
 with $c=100$.}
 \label{fig6}
\end{figure*}
\begin{figure*}[t!]
 \centering
 \includegraphics{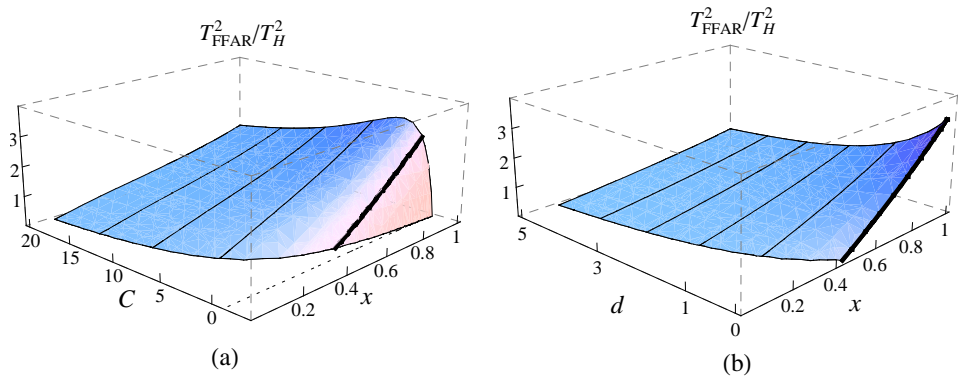}
 \caption{Freely falling temperature for the Schwarzschild-AdS black hole in massive gravity
 (a) by varying ${\cal C}$ with $d=0.1$, and (b) by varying $d$ with
 ${\cal C}=1$. Note that in (a) the dotted line on the $({\cal C},x)$ plane is for ${\cal C}=-1$. Thick curves
 represent freely falling temperatures for $({\cal C},~d)=(0,0.1),~(1,0)$.}
 \label{fig7}
\end{figure*}

The freely falling temperature (\ref{tffar2mschads}) can be
further simplified in the two interesting limit: at the spatial
infinity $r\rightarrow\infty$ ($x\rightarrow 0$) and at the event
horizon $r=r_H$ ($x=1$). At the spatial infinity of $x\rightarrow
0$, one obtains imaginary temperature as
 \be
 T^2_{\rm FFAR}=-\frac{1}{4\pi^2 l^2}<0,
 \ee
which is allowed for a geodesic observer who follows a
spacelike motion similar to the case of the Schwarzschild-AdS black
hole in the massless gravity
\cite{Deser:1997ri,Brynjolfsson:2008uc,Kim:2009ha}. On the other
hand, at the event horizon $x=1$, one obtains
 \be\label{tffar-event-horizon}
 T^2_{\rm FFAR}=\frac{c^2(1+{\cal C}+2d)}{4\pi^2 l^2}.
 \ee
It seems appropriate to comment that the freely falling
temperatures are finite at the event horizon, but become hotter
(colder) in the massive gravity with the condition $2d+{\cal C}>0$
($2d+{\cal C}<0$) than in the massless gravity with $d={\cal
C}=0$.

On the other hand, by taking the limit of $l^2\rightarrow\infty$
in Eq. (\ref{tffar2mschads}), one can find the freely falling
temperature of a Schwarzschild black hole in the massive gravity
of Eq. (\ref{tffarmsch}).
In the limit of ${\cal C}=0$ while keeping $d\neq 0$, the
temperature becomes
 \be
 T^2_{\rm FFAR}=\frac{x[(2d(1+2d)+(1+2d)^2)(1+x)+(1+2d)^2(x^2+x^3)]}
                 {16\pi^2r^2_H[2d+(1+2d)x]},
 \ee
which is drawn in Fig. 6(a) with a thick curve. In the limit of
$d=0$ with keeping ${\cal C}>-1$ except ${\cal C}=0$, we have
 \be
 T^2_{\rm FFAR}=\frac{(1+{\cal C})(1+x+x^2+x^3)}{16\pi^2r^2_H}>0,
 \ee
which behaves like the Schwarzschild-AdS black hole in the
massless gravity \cite{Brynjolfsson:2008uc,Kim:2009ha}. In Fig.
6(b), we have depicted a thick curve which is for $d=0$ with
${\cal C}=1$.  In the limit of both $d=0$ and ${\cal C}=-1$, we
have
 \be
 T^2_{\rm FFAR}=0.
 \ee
Moreover, in the case of ${\cal C}<-1$ with $d=0$, we
have
 \be
 T^2_{\rm FFAR}=\frac{(1+{\cal C})(1+x+x^2+x^3)}{16\pi^2r^2_H}<0,
 \ee
which is also allowed for a geodesic observer who follows a
spacelike motion for the case of the Schwarzschild black hole in
the massive gravity as expected. On the other hand, in the
massless case of both ${\cal C}=0$ and $d=0$, one can easily
obtain the squared freely falling temperature (\ref{tffar2-sch})
of the Schwarzschild black hole in the massless gravity
\cite{Brynjolfsson:2008uc,Kim:2009ha}.

\section{Discussion}

In summary, we have globally embedded the (3+1)-dimensional
Schwarzschild(-AdS) black hole in massless/massive gravity into
corresponding higher dimensional flat spacetimes. Making use of
the embedded coordinates, we have directly obtained the Hawking,
Unruh, and freely falling temperatures in a Schwarzschild(-AdS)
black hole in massive gravity and have shown that the Hawking
effect for a fiducial observer in a curved spacetime is equal to
the Unruh effect for a uniformly accelerated observer in a
higher-dimensionally embedded flat spacetime.

Moreover, we have shown that the GEMS embeddings of the
Schwarzschild-AdS black hole in massive gravity include not only
massive graviton effects but also AdS structures. Thus, in the
limit of $l^2\rightarrow\infty$, the GEMS embedding of the
Schwarzschild-AdS black hole in massive gravity is reduced to the
(5+2)-dimensional flat spacetime of the Schwarzschild black hole
in massive gravity, and furthermore in the massless limit of
${\cal C}\rightarrow 0$ and $R\rightarrow 0$, they are reduced to
the well-known (5+1)-dimensional flat spacetime. We have also
obtained their corresponding freely falling temperatures in these
limiting cases. Finally, we have found that freely falling
temperatures are finite at the event horizon, but become hotter
(colder) in massive gravity with the condition $2d+{\cal C}>0$
($2d+{\cal C}<0$) than in massless gravity with $d={\cal C}=0$.

Finally, it seems appropriate to comment on thermodynamic
stability briefly. The Hawking temperatures in Eqs.
(\ref{HTmSch}), (\ref{HT-mschads}) show that they are explicitly
modified by massive gravitons. First of all, as for the
Schwarzschild black hole in the massive gravity,  when ${\cal
C}>-1$, the Hawking temperature (\ref{HTmSch}) is exactly the same
form with the one in the massless gravity so that it would be
unstable. However, when ${\cal C}<-1$, it appears to have a stable
black hole. On the other hand, the Hawking temperature
(\ref{HT-mschads}) in the Schwarzschild-AdS black hole in the
massive gravity has a minimum temperature $T_m$ at $r_H=r_m$ as
seen in Fig. 4. As a result, when  ${\cal C}>-1$, a large black
hole which is defined by the one with $r_H>r_m$ would be stable,
while a small black hole with $r_H<r_m$ expects to be unstable.
Moreover, when ${\cal C}\le -1$, it would be stable since
$dT/dr_H>0$ on the whole range of $r_H$ in Fig. 4. Since the
thermodynamic stabilities differ due to the values of the mass
parameters, it would be interesting to study this topic more in
details for a later work as a further supplement of Ref.
\cite{Cai:2014znn}.

\acknowledgments{S. T. H. was supported by Basic Science Research Program through
the National Research Foundation of Korea funded by the Ministry of Education, NRF-2019R1I1A1A01058449.
Y. W. K. was supported by the National Research Foundation of Korea grant funded by the
Korea government, NRF-2017R1A2B4011702.}


\end{document}